\documentclass[10pt,letterpaper]{article}
\usepackage{opex3}
\usepackage{epstopdf}
\usepackage{amsmath}
\usepackage{subfigure}
\usepackage{cite}
\usepackage{graphicx,color}
\definecolor{mygreen}{rgb}{0,0.5,0} 
\definecolor{myrandomcolor}{rgb}{0.73,0.52,0.46} 
\definecolor{myblue}{rgb}{0,0,0.75} 
\definecolor{mymagenta}{cmyk}{0,1,0,0.12}

\newcommand{\gtext}[1]{{\color{mygreen}#1}}
\renewcommand{\gtext}[1]{{\color{mygreen}}}

\begin{document}
\title{Ultra-fast quantum randomness generation by accelerated phase diffusion in a pulsed laser diode \gtext{Ultra-fast quantum random number generation based on phase diffusion}}

\author{C.~Abell\'{a}n,$^{1, *}$~W.~Amaya,$^{1}$~M.~Jofre,$^{1}$~M.~Curty,$^{2}$~A.~Ac\'{i}n,$^{1,3}$~J.~Capmany,$^{4}$ V.~Pruneri,$^{1,3}$ and~M.~W.~Mitchell,$^{1,3}$}
\address{$^{1}$ ICFO-Institut de Ciencies Fotoniques, Castelldefels, 
E-08860 Barcelona, Spain\\
$^{2}$ EI Telecomunicaci\'{o}n, Dept. of Signal Theory and Communications, University of Vigo, E-36310 Vigo, Spain\\
$^{3}$ ICREA-Instituci\'{o} Catalana de Recerca i Estudis Avan\c{c}ats, 
E-08010 Barcelona, Spain\\
$^{4}$ ITEAM-Research Institute, Universidad Polit\'{e}cnica de Valencia, E-46022 Valencia, Spain}
\email{*carlos.abellan@icfo.es} 

\begin{abstract*} 
We demonstrate a high bit-rate quantum random number generator by interferometric detection of phase diffusion in a gain-switched DFB laser diode.  Gain switching at few-GHz frequencies produces a train of bright pulses with nearly equal amplitudes and random phases.  An unbalanced Mach-Zehnder interferometer is used to interfere subsequent pulses and thereby generate strong random-amplitude pulses, which are detected and digitized to produce a high-rate random bit string.  Using established models of semiconductor laser field dynamics, we predict a regime of high visibility interference and nearly complete vacuum-fluctuation-induced phase diffusion between pulses.  These are confirmed by measurement of pulse power statistics at the output of the interferometer.  Using a 5.825 GHz excitation rate and 14-bit digitization, we observe {43} Gbps quantum randomness generation.
\end{abstract*}
%


\section{Introduction}

Random number generators (RNG) have been extensively studied by different groups because of the wide spectrum of their applications: secure communications \cite{Tajima2007}, stochastic simulation \cite{Cai2007} and gambling \cite{Hall1997}, among others.

Deterministic algorithms known as pseudo-random number generators can rapidly generate bit sequences with long repetition lengths, and are often used as a substitute for random numbers.  Physical RNGs take as input data from a physical process believed to be random, for example electronic and thermal noise \cite{Petrie2000}, chaotic semiconductor lasers \cite{KanterNP2010} and amplified spontaneous emission signals \cite{Argyris2012}. A subset of physical RNGs, known as quantum RNGs (QRNGs) use a physical process with randomness derived from quantum processes.  Processes used include radioactive decay \cite{YoshizawaJJSCS1999}, two-path splitting of single photons \cite{Jennewein2000}, photon number path entanglement \cite{Kwon2009}, amplified spontaneous emission \cite{Williams2010}, measurement of the phase noise of a laser \cite{Guo2010, Jofre2011, Xu2012},  photon arrival time \cite{WahlAPL2011}, and vacuum-seeded bistable processes \cite{MarandiOE2012}.  These physical RNGs and QRNGs show a trade-off between speed of generation and surety of the random bits generated: chaotic and ASE sources \cite{KanterNP2010,Argyris2012,LiOL2012,WangOE2013} reach hundreds of Gbps using signals that include contributions from both random and in-principle predictable sources, e.g. detector noise.  In contrast, QRNGs can guarantee the quantum origin and thus the randomness of the signal \cite{Jofre2011,WahlAPL2011,PironioN2010}, although to date at lower bit rates.  

Recently, Jofre, et al. \cite{Jofre2011} and Xu, et al. \cite{Xu2012} have reported quantum random number generation using phase diffusion in semiconductor lasers. In \cite{Jofre2011} a QRNG rate up to 1.1 Gbps was demonstrated and in \cite{Xu2012}, 6 Gbps QRNG rate. The main difference is that Jofre et al. strongly modulate the laser diode, taking it below and above threshold, while Xu et al. directly detect phase fluctuations in an above-threshold continuous wave regime. For the same mean power, pulsing accelerates the phase diffusion rate, which is proportional to the spontaneous emission rate over the intra-cavity photon number.   Here we report improvements in both speed and surety in the Jofre et al. method.  We {achieve} 43 Gbps by using a higher bandwidth laser diode and a faster modulating system. {Also, we demonstrate a new method to lower-bound the quantum contribution to the randomness of the observed pulse distribution.}

\section{System configuration}

\begin{figure}[*htp]
\begin{center}
\includegraphics[scale=0.5]{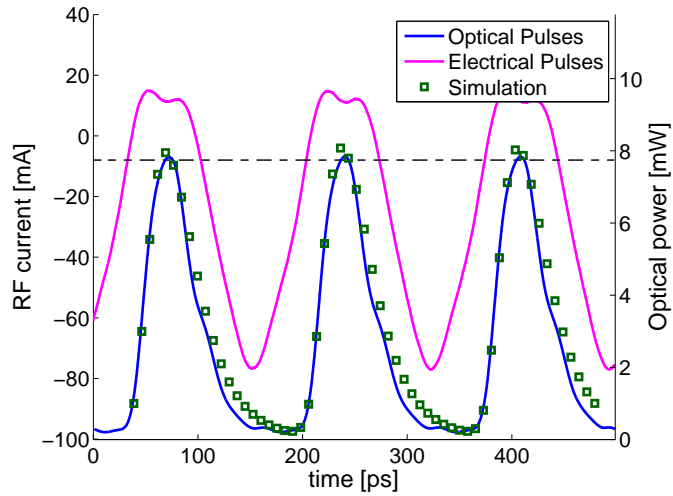}
\caption{Electrical and optical pulse trains. Magenta, (upper trace): electrical current drive applied to the laser, with PRF of 172 ps. Blue, (lower trace): photo-detected optical pulses of 85 ps time width and 7.65 mW peak power and (black, dashed line)  9 mA LD current threshold. Simulation is a conservative fitting of the rate equations such that the predicted detected output power vs. time is always larger than the  observed output power vs. time.}
\label{fig:pulsos}
\end{center}
\end{figure}

{A 1550 nm distributed feedback (DFB) laser diode (LD) with 10 Gbps modulation bandwidth} is driven through its DC input with a constant current of 15 mA, and through its AC-coupled RF input with a waveform modulated at 5.825 GHz.  These two sources, combined with a bias-tee and recorded on an oscilloscope, are shown in Fig. \ref{fig:pulsos}.  Because the RF drive amplitude exceeds the bias, the diode is reverse-biased for about 40\% of the cycle, creating strong attenuation within the material between periods of high gain.  The resulting optical output consists of  pulses of $\sim$ 85 ps  width and $\sim$ 7.65 mW peak power.  To avoid back reflections into the oscillator cavity, the LD has an internal 30 dB optical isolator.

\newcommand{\laserphase}{\theta}
\newcommand{\armphase}{\phi}
\newcommand{\RF}{{\rm RF}}

\begin{figure}[htp]
\begin{center}
\includegraphics[scale=0.7]{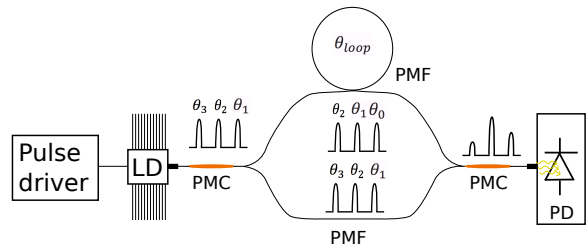}
\caption{ Unbalanced Mach-Zehnder interferometer (U-MZI). Phase-randomized coherent optical pulses interfering at the output of the U-MZI produce random intensities. (Pulse driver) denotes the electrical pulse generator that directly modulates the laser, (LD) laser diode, (PMC) polarization maintaining coupler, (PMF) polarization maintaining fiber, ($\laserphase_{0-3}$) optical phases of different consecutive pulses, ($\theta_{loop}$) phase introduced by the delay line and (PD) a fast photodetector. }
\label{fig:blockdiagram}
\end{center}
\end{figure}

The linearly polarized optical pulses pass through an unbalanced Mach-Zehnder interferometer (U-MZI) composed of two 50/50 polarization maintaining couplers (PMC) with arm-length difference related to the pulse repetition frequency (PRF), see Fig. \ref{fig:blockdiagram}. The 3.55 cm path delay length is designed to interfere pulses at a PRF of 5.825 GHz. The optical output is detected by a 14-bit oscilloscope (DSA8200 with module 80C02), with 12.5 GHz bandwidth and triggered by the system clock reference, taken from the electrical pulse generator (Anritsu MP1800A).

\newcommand{\rbs}{\varepsilon_{\rm 12}^{(1)}}
\newcommand{\tbs}{\varepsilon_{\rm 11}^{(1)}}
\newcommand{\rbsp}{\varepsilon_{\rm 21}^{(2)}}
\newcommand{\tbsp}{\varepsilon_{\rm 11}^{(2)}}
\newcommand{\tbspgen}{\varepsilon_{ij}^{(k)}}
\newcommand{\unoise}{u_{\rm noise}}
\newcommand{\var}{{\rm var}}

\section{Principle of operation}
The field detected at the output of the interferometer is
\begin{equation}
\label{Eq:Eout}
{\cal{E}}^{\rm (out)}(t)={\tbs \tbsp} {\cal{E}}^{\rm (laser)}(t-t_1) + {\rbs \rbsp}  {\cal{E}}^{\rm (laser)}(t-t_2),
\end{equation}
where $\tbspgen$ is the field transmission coefficient of the $k$th PMC from input port $i$ to output port $j$,  ${\cal{E}}^{\rm (laser)}$ is the field emitted by the laser and $t_{1,2}$ are the delays caused by arms 1,2 of the U-MZI, respectively.  For convenience, we  write ${\cal{E}}^{\rm (laser)}(t) = A(t) \exp[-i \omega t]  \exp[i \laserphase_j ]$ where $\omega$ is the central frequency of the laser emission and $\laserphase_j$ is the phase of the $j$th pulse.  Defining the powers $u^{\rm (out)}(t) \equiv  |{\cal{E}}^{\rm (out)}(t)|^2$ ,  $u_{1}(t)  \equiv |\tbs \tbsp {\cal{E}}^{\rm (laser)}(t-t_{1})|^2$ and $u_{2}(t)  \equiv |\rbs \rbsp {\cal{E}}^{\rm (laser)}(t-t_{2})|^2$ we have the observed signal
\begin{equation}
u^{\rm (out)}(t)= u_1(t) + u_2(t) + 2|g^{(1)}(t)|\sqrt{u_1(t) u_2(t)}\cos (\laserphase_j - \laserphase_{j-1} + \Delta\phi) + \unoise,
\label{Eq:uout}
\end{equation}
where $\Delta \phi \equiv \omega (t_1-t_2)$ is the relative phase delay of the two arms, 
$g^{(1)}(t) \equiv \langle {\cal{E}}^{*\rm (laser)}(t-t_1) {\cal{E}}^{\rm (laser)}(t-t_2) \rangle/\sqrt{\langle|{\cal{E}}^{\rm (laser)}(t-t_1) |^2 \rangle \langle |{\cal{E}}^{\rm (laser)}(t-t_2)|^2 \rangle}$ is the degree of first-order coherence (the visibility) and here $\langle \cdot \rangle$ indicates an average over the response time implied by the finite bandwidth of the detector and recording electronics. $\unoise$ is noise contribution from the detection and digitization electronics.  Fluctuation of $\Delta \phi$ due to changes in $t_1-t_2$, e.g. stretching of the fiber, were measured by above-threshold continous-wave operation of the laser, which showed RMS phase fluctuations, between points separated by 1/PRF, of $\le 2 \times 10^{-7}$ rad, which is negligible on the scale of the quantum fluctuations described below.

\section{Phase diffusion}
\label{Sec:PD}

The experiment is performed under gain-switching conditions, and the field within the cavity alternately experiences two working regimes: far above threshold and far below threshold. During the former, stimulated emission dominates and coherent optical pulses are emitted. During the latter, the intracavity field is strongly attenuated and significant phase diffusion occurs. The stochastic nature of this process has been studied in detail by C. Henry and G. Agrawal in \cite{Henry1982, HenryJLT1986, Agrawal1990}, where an expression for the average phase diffusion $\langle \Delta \laserphase^2 (t) \rangle$ is derived under various conditions from the Langevin equation
\begin{equation}
\label{Eq:phase}
\dot{\laserphase} = \frac{\alpha}{2}G_N(n - n_{th}) - \frac{\beta_{\rm SE}}{2}\frac{G_N (n - n_0) p}{1 + \sqrt{1+p}} + F_{\laserphase}(t),
\end{equation}
where $F_{\laserphase}$ is a Langevin force responsible for the phase diffusion process, $\alpha = 5.4$ is the linewidth enhancement factor first derived by C. Henry \cite{Henry1982} and all other parameters are described below. The linear approximation proposed in \cite{HenryJLT1986}

\begin{equation}
\label{Eq:PhaseVariance}
\frac{d}{dt}\langle \Delta \laserphase (t)^2 \rangle = \frac{R_{sp}}{2 s}(1 + \alpha^2),
\end{equation}
gives a lower bound on the phase diffusion in an interval of time $t$.  Here $R_{sp}$ is the  spontaneous emission rate, which depends {(see next paragraph)} on the number of carriers $n$, and $s$ is the number of photons in the cavity \cite{Henry1982}. Hence, the dynamical behavior of $R_{sp}$ and $s$ are described by differential equations, given in \cite{Agrawal1990}, Eqs. (13)--(17), which couple $s$ to $n$. 
\begin{align}
\label{Eq:difphotons}\dot{s} &= G_N \Big(\frac{n-n_0}{\sqrt{1 + s/s_{sat}}} - (n_{th} -n_0) \Big) s + R_{sp},\\
\label{Eq:difcarriers}\dot{n} &= I/q - \gamma_e n - G_N \frac{n - n_0}{\sqrt{1 + s/s_{sat}}} s,
\end{align} 
{where $G_N$ is the gain per carrier, $n_0$ is the number of carriers at transparency, $n_{th}$ is the number of carriers at threshold, $s_{sat}$ is the saturation photon number, $\gamma_e$ is the carrier recombination lifetime $= 1/\tau_e \approx 1 \times 10^{9}$, $I$ the injection current and $q$ is the electronic charge.  To give values to these parameters, we note that $G_N = \gamma/(n_{th} - n_0)$, where $\gamma = c (\alpha_m+\alpha_s)/\bar{n}$ is the cavity decay rate where $c$ is the speed of light in vacuum, $\bar{n} = 4.33$ is the effective refractive index,  $\alpha_s = 4.5\; cm^{-1}$ describes losses due to scattering and absorption and $\alpha_m \approx 1.4/L$ is the cavity escape loss, where $L$ is the cavity length.  Because $s_{sat}, L$ and $n_0$ are unknown to us, we choose them by fitting to the observed pulse-shape (see below).  $R_{sp}$ can be written $R_{sp} = n \gamma_e  R_0$, where {$R_0 = K_{\rm tot} \Gamma_{\rm conf} \beta_{\rm SE}$} is a constant containing $\Gamma_{\rm conf}$ the ``confinement factor'', $\beta_{\rm SE}$ the fraction of spontaneous emission coupled to the lasing mode, and $K_{\rm tot}$ the ``total enhancement factor'' in a DFB LD, i.e. the enhancement in the fraction of spontaneous emission actually coupled to the lasing mode because of the geometry.  We also do not have access to these parameters, but their combined effect is measurable using the steady-state power (see below).  

The solution of the rate equations, i.e. $s(t)$ and $n(t)$, permits the calculation of $R_{sp}$ and therefore the dynamical evolution of the average phase diffusion $\langle \Delta \laserphase^2\rangle$ in Eq. (\ref{Eq:PhaseVariance}). In order to fit a solution of these equations to the measurements on our DFB LD, we first need to estimate the parameters $s_{sat}$, $L$, $n_{th}$, $G_N$, $n_0$ and $R_0$ (these are not all independent). Also, the solution is low pass filtered with a single-pole recursive filter with a time constant of $\tau_{f} = 0.35/BW_{osc}$, where $BW_{osc} = 12.5$ GHz is the bandwidth of the oscilloscope.

A recursive method is used to extract  these parameters:
\begin{enumerate}
	\item  We set $s_{sat}$ to a reasonable value, and $L$ to one of $100, 200, 500$ or $1000\,\mu$m.
	\item  We solve the steady-state solution of the differential equations, Eq. (\ref{Eq:SSp}) and Eq. (\ref{Eq:SSn}), near threshold, i.e., with $n \approx n_{th}$.  The DFB LD emits $s_{th'} = 0.3$ mW when it is biased with a constant current of $I_{th'} = 10$ mA. (The subscript $_{th'}$ refers to steady-state solution slightly above threshold).  Eqs. (\ref{Eq:difphotons}) and (\ref{Eq:difcarriers}) then take the form 
		\begin{align}
			\label{Eq:SSp}\gamma \Big(\frac{1}{\sqrt{1 + s_{th'}/s_{sat}}} -1\Big) s_{th'} + R_0 \gamma_e  n_{th} &=0,\\
			\label{Eq:SSn} \frac{I}{q} - \gamma_e n_{th} - \gamma s_{th'} \frac{1}{\sqrt{1+s_{th'}/s_{sat}}}&=0,
		\end{align}
from which we  extract $n_{th}$ and $R_0$. 	
	\item   We choose the maximum $G_N$, which mainly controls the speed of the dynamics, such that the predicted detected output power vs. time is always larger than the observed output power vs. time (from Fig. \ref{fig:pulsos}). Note that setting $G_N$ directly specifies $n_0$, via $G_N = \gamma/(n_{th} - n_0)$. 
	\item  We repeat steps (1)--(3) to find the values of $s_{sat}$  and $L$ that minimize the rms deviation of the simulation from the observed power curves.
\end{enumerate}

After these steps, $s_{sat} = 7.7\times 10^5$,  $L = 500\,\mu$m, $n_{th} = 5.62\times 10^7$, $R_0 = 8.8 \times 10^{-4}$, $G_N = 2.3 \times 10^4$ and $n_0 = 3.46\times 10^7$. The result is shown in Fig. \ref{fig:pulsos}. Note that phase diffusion is larger for low optical intensities. Hence, the fitting is conservative because the simulated optical power is always larger than the measurement.
}

Finally, Eq. (\ref{Eq:PhaseVariance}) is numerically integrated to find  $\langle \Delta \laserphase^2 \rangle > (9.45 {\rm ~ rad})^2$. For all practical purposes this value describes a full randomization.  We note that $\cos \theta$ is unchanged under $\laserphase \rightarrow \laserphase + 2 \pi n$ ($n$ integer) and under $\laserphase \rightarrow -\laserphase$.  If $\laserphase$ is described by a Gaussian PDF $G(\theta)$ with width ${\langle \Delta \laserphase^2 \rangle}$, then an equivalent distribution, for the purposes of the cosine, is $G_\pi(\laserphase) \equiv \sum_{s = \pm 1} \sum_{n=-\infty}^\infty G(s \laserphase + 2 \pi n)$ for $0 \le \laserphase < \pi$ and $G_\pi(\laserphase) \equiv 0$ otherwise.  Already with ${\langle \Delta \laserphase^2 \rangle} = (2 \pi)^2$,  $G_\pi$ approximates a uniform distribution on $[0,\pi)$ with a fractional error below $10^{-8}$.

\begin{figure}[*htp]
\begin{center}
\subfigure[Histograms]{\label{fig:Histograms}\includegraphics[width=0.48\linewidth]{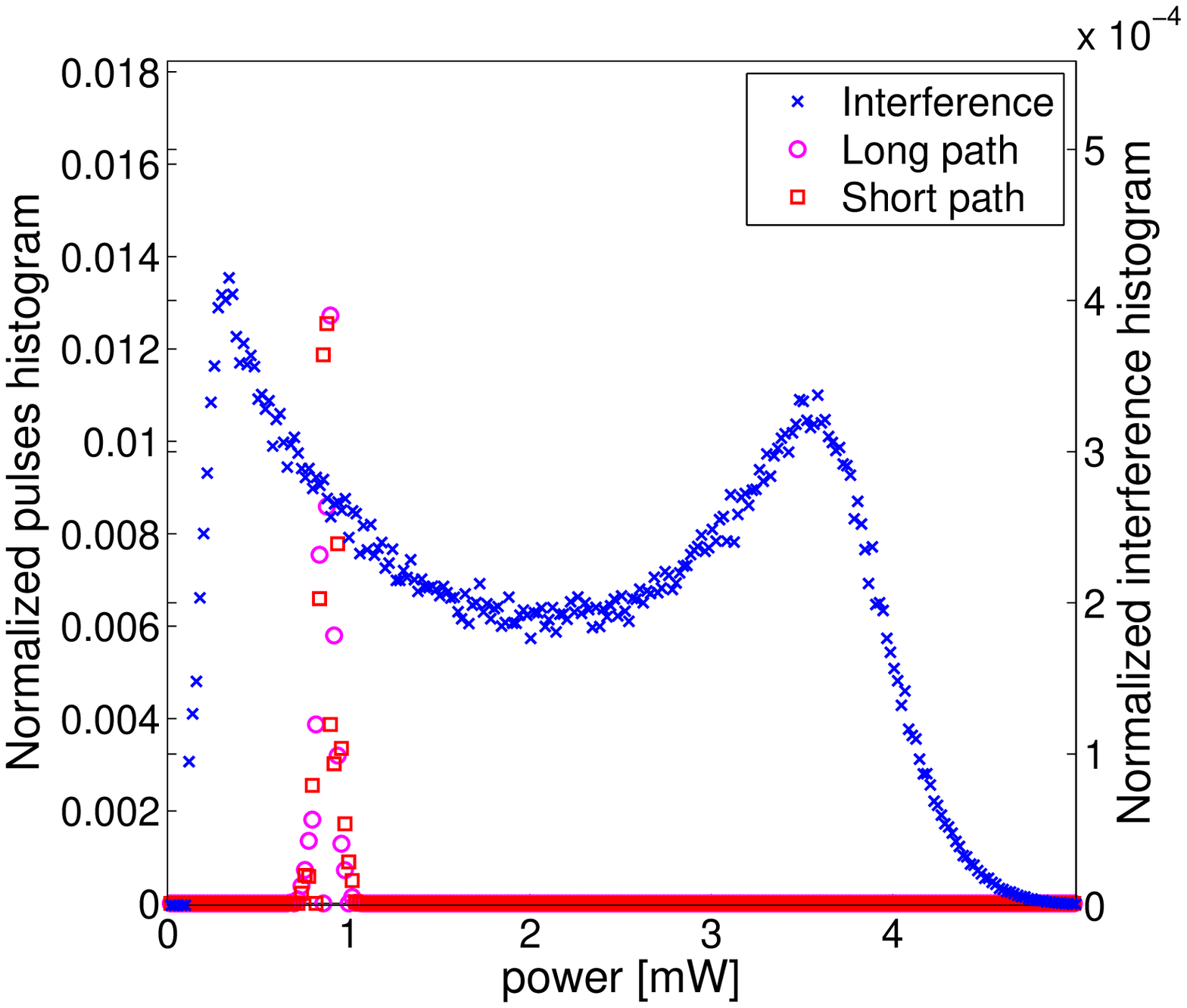}}
\subfigure[Raw data correlation]{\label{fig:rawCorrelation}\includegraphics[width=0.48\linewidth]{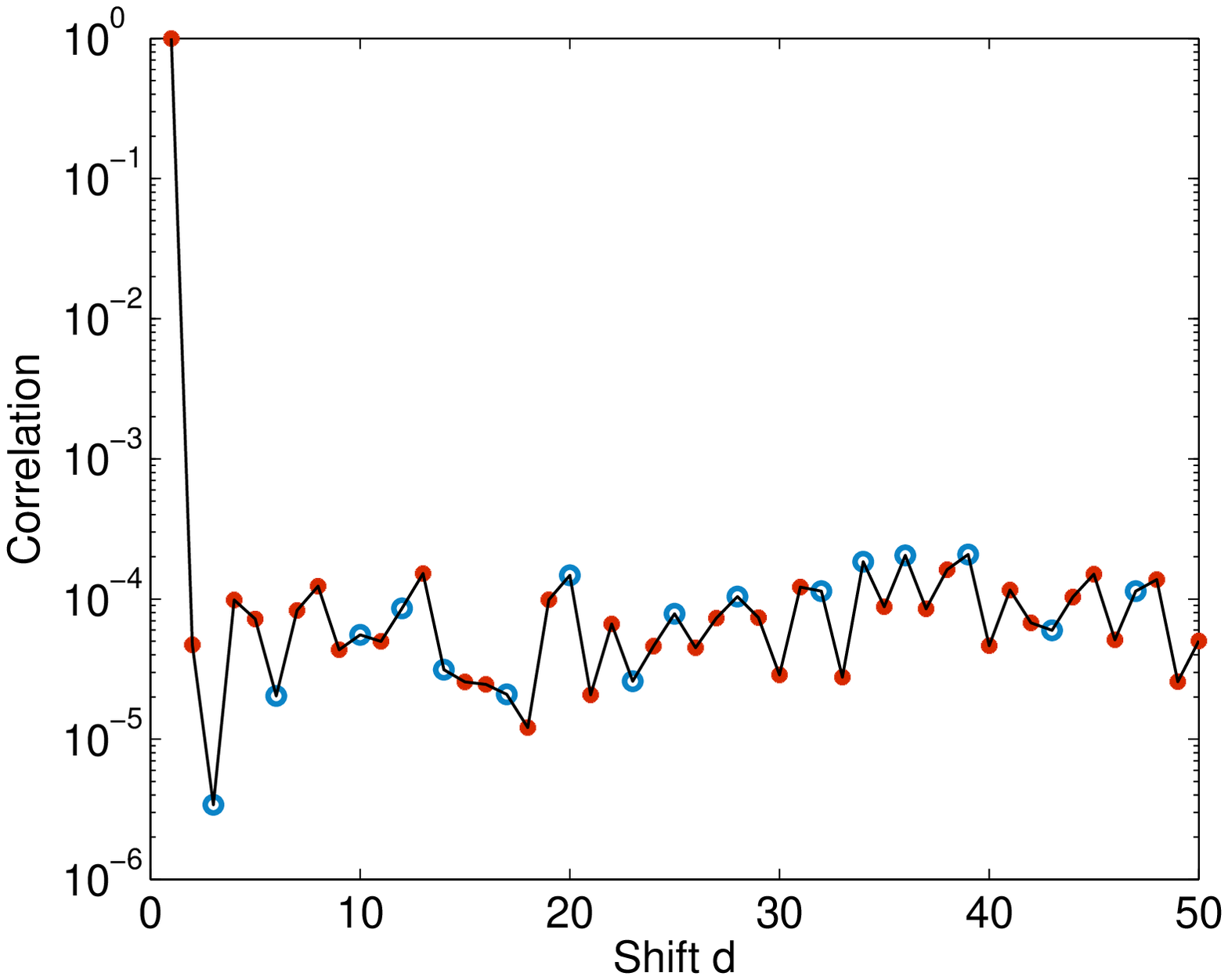}}
\caption{In Fig. \ref{fig:Histograms}, input power distributions (left y axis) and the resultant output power distribution (right y axis). The visibility achieved for the interferometer is 0.9. The power distribution has clearly widened due to the random phase generated by amplified spontaneous emission (ASE). In Fig. \ref{fig:rawCorrelation}, normalized correlation of 50 subsequent sampled pulses. The autocorrelation has been evaluated with $120 \times 10^6$ 14-bit samples, but just the first 50 terms are shown. As  expected, it follows a delta-function like behaviour indicating the random nature of the process.}
\label{fig:RawData}
\end{center}
\end{figure}

\section{\textbf{Raw data analysis}}
A RNG should produce an uncorrelated and uniformly distributed string of numbers. The uncorrelation avoids the capability to predict future numbers while the uniformity ensures the same apparition probability to all of them. Sampling the interfered field, an uncorrelated but not equally distributed sequence of numbers is achieved. The nearly complete phase diffusion is expected to provide uncorrelated pulses, while the uniformity will be achieved using a randomness extractor over the raw data.

For statistical testing, $120 \times 10^{6}$ pulses were recorded on a fast 14-bit sampling oscilloscope.  From each pulse a single sample was recorded, taken $\sim$ 13 ps after the pulse peak, to produce $120 \times 10^{6}$ 14-bit numbers. This delay was chosen experimentally to give a broad distribution in Fig. \ref{fig:Histograms}. We observe experimentally that the distribution takes on the expected shape after power saturation occurs, i.e. after the pulse peak, presumably due to damping of transients associated with the rapid turn-on. These data were taken over a five day period as a test of stability and repeatability.
 
The power from each branch of the U-MZI, as measured at the output, are $0.97$ mW and $0.9$ mW on average, with a standard deviation of $\sim 45 \, \mu W$. This deviation around the mean arises from background noise and spontaneous emission events, producing narrow distributions for the interfering pulses. In contrast, due to interference of equal power optical pulses with random phases, the output power distribution is expected to broaden. The fact that the optical pulses have a predictable power and true random phase implies that when interfering these fields, random input phases of quantum origin are converted into macroscopically measurable random output intensities. This behaviour is shown in Fig. \ref{fig:Histograms}.

We evaluate the normalized autocorrelation function  to confirm  the intuition that the process is intrinsically uncorrelated. As seen in Fig. \ref{fig:rawCorrelation}, the non-shifted correlation values are at the 40 dB level.

\newcommand{\pdf}{{PDF }}

\section{\textbf{Randomness extraction and post-processing analysis}}
Randomness extractors transform non-uniformly-distributed sequences into uniformly-distributed sequences \cite{Nisan1999} at the cost of losing a fraction of the bits. Given a random variable $X$ distributed according to the distribution $P[X=x_i]$, the 
  \emph{min-entropy} 
 \begin{equation}
\label{Eq:minentropy}
H_{\infty}(X) \equiv -\log_2 \Big( \max_{\substack{\forall \; x_i}} P[X = x_i] \Big),
\end{equation}
quantifies the amount of extractable randomness, in the sense that from a sequence $\{X_1, \ldots, X_N\}$ with $N \gg 1$, a uniform random bit sequence of length $N H_{\infty}(X)$ can be extracted.  For our digitized input with $b$ bits of resolution, we define the reduction factor $\RF \equiv b/H_{\infty}(X)$.  

The detected signal given by Eq. (\ref{Eq:uout}) contains a large random component through the last term $\sqrt{u_1 u_2}\cos \phi$, where $\phi$ is the random phase due to phase diffusion.  If this were the only fluctuating term, the observed signal would obey an arcsine distribution, the distribution of the cosine of a uniformly-distributed phase.  Other fluctuating contributions to the signal, including photo-detection noise, laser current fluctuations, and digitization errors, broaden and smooth this ideal distribution to give the observed distribution shown in Fig. \ref{fig:RawData}.  These other noise sources are not guaranteed to be random.  Fortunately, the randomness extraction step will remove any non-random effects of these other noise sources, provided that the reduction factor is chosen using the min-entropy {of the quantum contribution}.  In what follows, we show that this quantum contribution can be determined from second order statistics of the measured observable and noise sources, plus our knowledge about the distribution of the phase.

\subsection{\textbf{Procedure}}
The interfering process described in Eq. (\ref{Eq:uout}) contains four random variables: $X_1 = u_1$, $X_2 = u_2$, $X_3 = \sqrt{u_1 u_2}\cos \theta$ and $X_4 = \unoise$, where $X_3$ is  the product of three random variables $\sqrt{u_1}$, $\sqrt{u_2}$ and $\cos \theta$. Assuming that $u_1$, $u_2$, $\cos \theta$ and $\unoise$ are independent, does not imply that the four random variables $X_1, X_2, X_3, X_4$ are independent. In fact, $\sqrt{u_1 u_2}\cos \theta$ depends on $u_1$ and $u_2$. Nevertheless, the fact that $\cos \theta$ is equally likely to be positive or negative implies that the four random variables are uncorrelated. 

The arcsine probability distribution function, its mean and its variance are given by:
\begin{equation}
\pdf(x) = \frac{1}{\pi \sqrt{(x-a)(b-x)}}; \hspace{5mm} \mu_{x} = \frac{a + b}{2}; \hspace{5mm} \text{var}[x] = \frac{1}{8}(b - a)^2.
\label{Eq:Arcsine}
\end{equation}

In order to estimate the visibility from the observed distributions of $u_1, u_2$ and $u^{(\rm out)}$, the variance at both sides of Eq. (\ref{Eq:uout}) is calculated. Assuming uncorrelation:
\begin{equation}
\label{Eq:VarOutI}
\var(u^{(out)}) = \var(u_1) + \var(u_2) + \var(\unoise)+ 4 |g(t_{loop})|^2 \var(\sqrt{u_1 u_2}\cos \laserphase).
\end{equation}
From laser physics, we know that the intra-cavity field describes a diffusion process and therefore $\laserphase$ is gaussianly distributed. Also, using measures and well-established models of semiconductor DFB lasers, an standard deviation larger than $\pi$ was estimated in Section \ref{Sec:PD}.  As described above, this justifies considering $\laserphase$ to be uniformly distributed on $[0, \pi)$, and thus that $\cos \theta$ is described by an arcsine distribution. From Eq. (\ref{Eq:Arcsine}) and Eq. (\ref{Eq:VarOutI}), it then follows that

\begin{equation}
|g(t_{loop})|^2 \approx \frac{\var(u^{(out)}) - \var(u_1) - \var(u_2) - \var(\unoise)}{2E[{\sqrt{u_1}}]^2E[\sqrt{u_2}]^2},
\label{eq:gExtraction}
\end{equation}
where $E[\cdot]$ indicates the mean value. 

From the data shown in Fig. \ref{fig:RawData}, we obtain $\text{var}(u^{(out)}) = 1.4$ {mW}$^2$, $\text{var}(u_1) = 2.0\times 10^{-3} $ {mW}$^2$, $\text{var}(u_2) = 2.1\times 10^{-3} $ {mW}$^2$, $E[\sqrt{u_1}]^2 = {0.97 {~\rm mW}}$, and $E[\sqrt{u_2}]^2 = {0.90 {~\rm mW}}$.  From measurements with the laser off, we find the contribution of detection and digitization noise $\text{var}(\unoise) = 1.45\times 10^{-4} $ {mW}$^2$.  Applying Eq. (\ref{eq:gExtraction})  we find $|g(t_{loop})| = 0.90$.

\subsection{\textbf{Min-entropy estimation}}
\newcommand{\umin}{{u_{\rm min}}}
\newcommand{\umax}{{u_{\rm max}}}

The distribution produced by the random phase, i.e. without considering undesired random effects, is given by an arcsine distribution between $\umin$ and $\umax$, where
  \begin{eqnarray}
   \umin &\equiv& E\left[ u_1 + u_2 - 2|g(t_{loop})| \sqrt{u_1u_2} \right], \\
    \umax &\equiv& E\left[ u_1 + u_2 + 2|g(t_{loop})| \sqrt{u_1u_2} \right].
  \end{eqnarray}
Because the arcsine distribution is peaked at $x=\umin$, and because the min-entropy only depends on the weight of the most probable event, the min-entropy of the digitized arcsine distribution only depends on the probability of the first bin. 
If $A_{\textsc{adc}}$ is the dynamic range of the analog-to-digital converter (ADC), $b$ its resolution in bits so that $\Delta u = A_{\textsc{adc}}/2^b$ is the bin size, the probability mass of the first bin is
\begin{eqnarray}
P[X = x_1] &=& \frac{1}{\pi}\int_\umin^{\umin+\Delta u} \frac{1}{\sqrt{(u-\umin) (\umax-u)}} du \nonumber \\
& = &  \frac{2}{\pi}\arcsin \sqrt{\frac{A_{\textsc{adc}}}{2^b (\umax-\umin)}}.
\end{eqnarray}
Using the fact that $\arcsin x \approx x$ for $x$ small, it can easily be seen that the probability of the first bin decreases exponentially with $b$ and therefore the min-entropy, given by Eq. (\ref{Eq:minEntropy}), increases linearly with the resolution of the digitization. 
\begin{align}
\label{Eq:minEntropy}
H_{\infty}(X)&\approx  \frac{b}{2} - \frac{1}{2} \log_2 \Big( {\frac{4 A_{\textsc{adc}}}{\pi^2 (\umax-\umin)}}\Big).
\end{align}
In the experiment, where $A_{\textsc{adc}} = 5 $ mW, $b=14$ bits,  $\umax-\umin= 4 |g(t_{loop})| \sqrt{E[{u_1]E[u_2]}} =  3.34$ mW, the min entropy is ~7.33 bits.  With the PRF of 5.825 Gpulses/s, this implies a quantum randomness generation rate of  43  { Gbps}.

\begin{figure}[*htp]
\centering
\subfigure[Auto-correlation of hashed data]{\label{fig:hashedCorrelation}\includegraphics[scale=0.4]{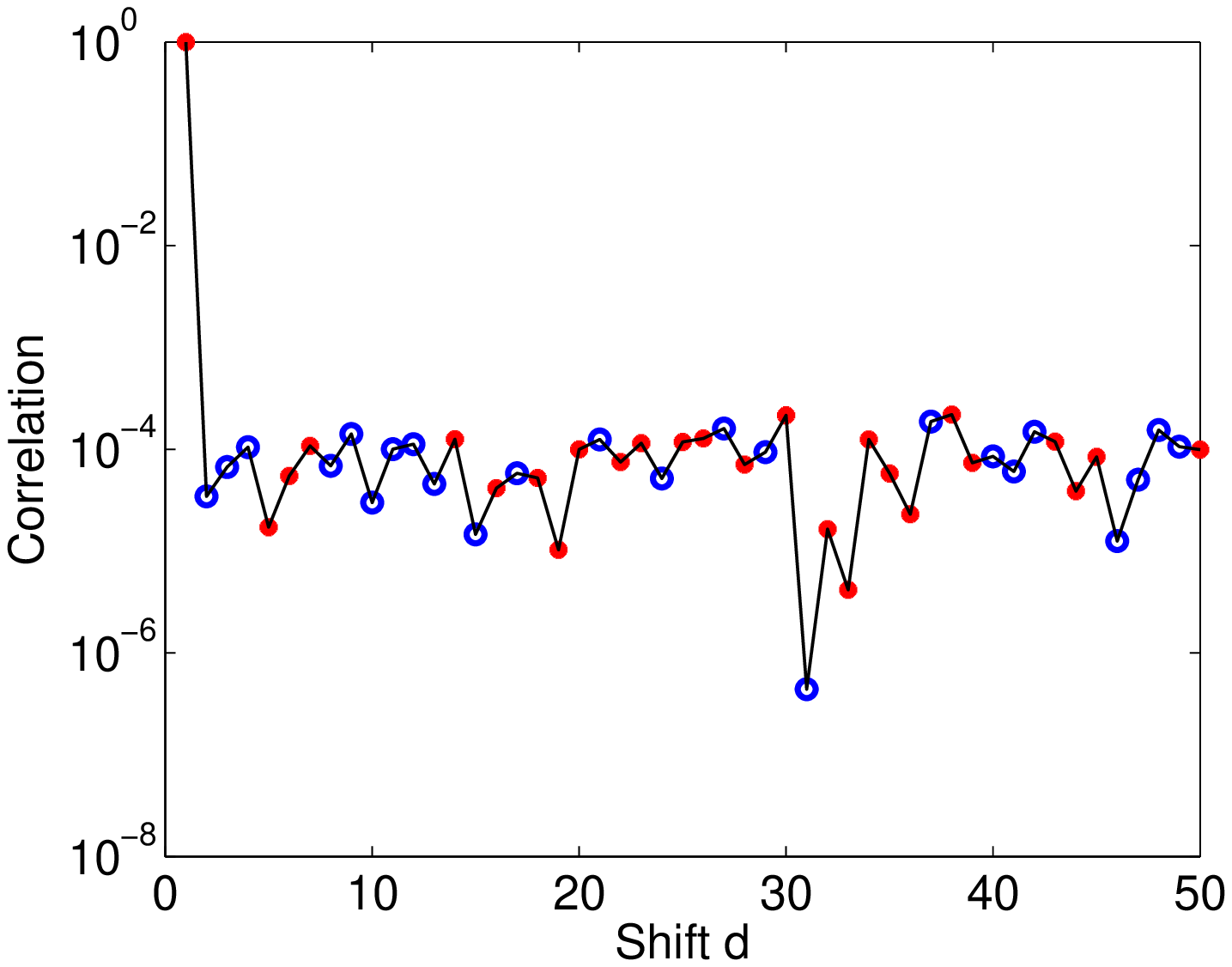}}
\subfigure[Deviation from uniform distribution]{\label{fig:deviation}\includegraphics[scale=0.32]{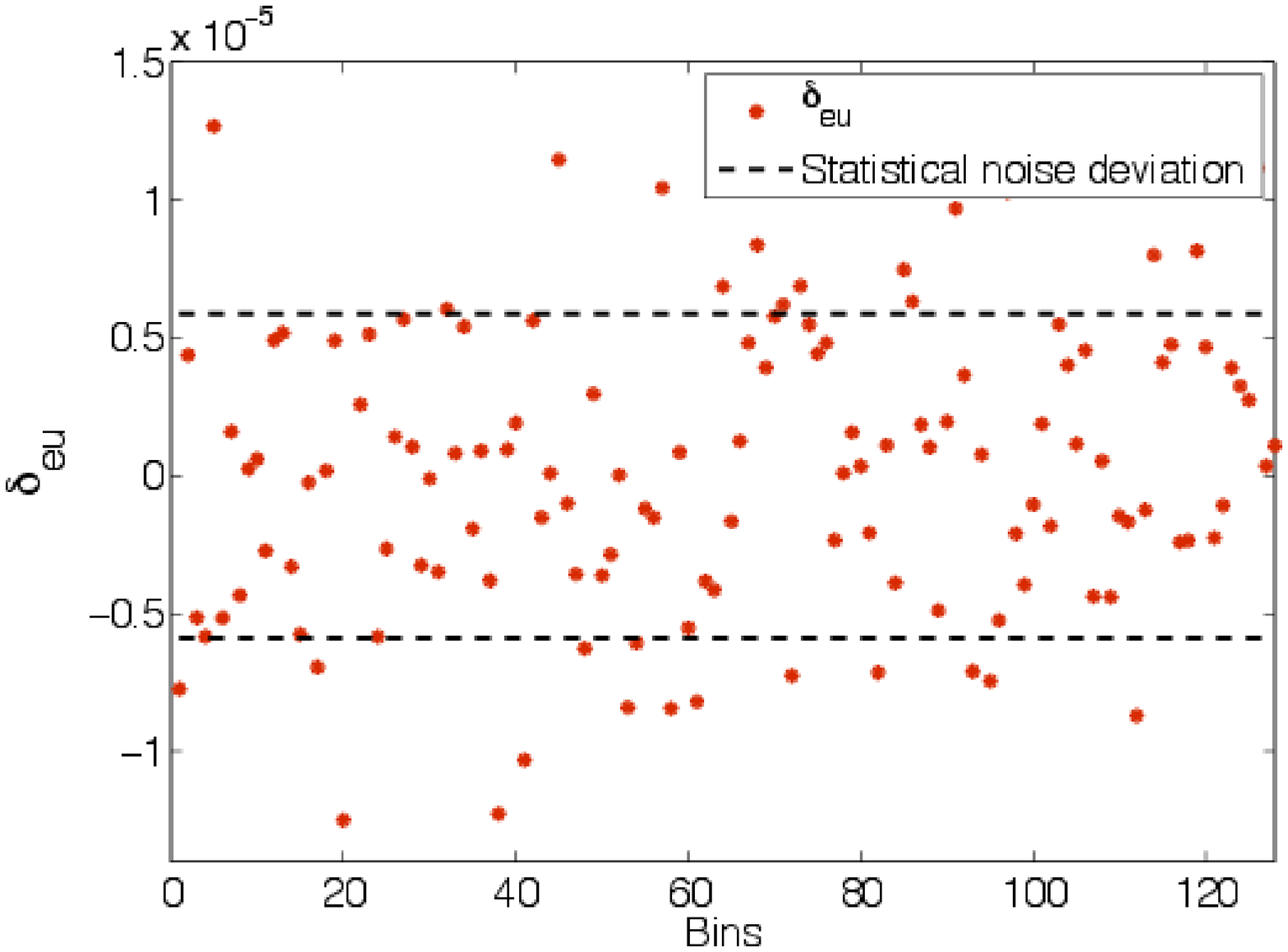}}
\caption[Hashed autocorrelation functions]{{Statistical characterization of $125 \times 10^6$ 7-bit numbers produced by hashing the experimental data.  (a)  autocorrelation of the hashed data.  Autocorrelation at the -40 dB level is seen, corresponding to the expected statistical variation. 
 
(b) deviation from the ideal 7-bit uniform distribution.  Dashed lines show plus/minus 1 sigma expected variation. }} \label{Fig:hashedCorr}
\end{figure}

In order to extract the randomness arising from amplification of the vacuum field, a hash function with a reduction factor $\RF = 14/7.33 \approx 1.9$ is applied to the raw data, reducing $120 \times 10^6$ 14-bit numbers to $125 \times 10^6$ 7-bit numbers. The hashing is performed by the Whirlpool hash function \cite{WhirlpoolURL}.  As seen in Fig. \ref{Fig:hashedCorr}, correlations within the resulting data are at the -40 dB level, which coincides with the statistical uncertainty given the quantity of output numbers.  To check uniformity, we compute the frequencies of the 7-bit numbers.  Fig. \ref{fig:deviation} shows the deviation from the ideal 7-bit uniform distribution $\delta_{eu} = P_{i} - 1/128$, where $P_i$ is the probability of the $i$-th bin.  No statistically significant deviation from uniformity is observed.

\newcommand{\incompletegamma}{\Gamma}

\begin{figure}[!h]
\centering
\subfigure[Proportion]{\label{Fig:proportion}\includegraphics[scale=0.4]{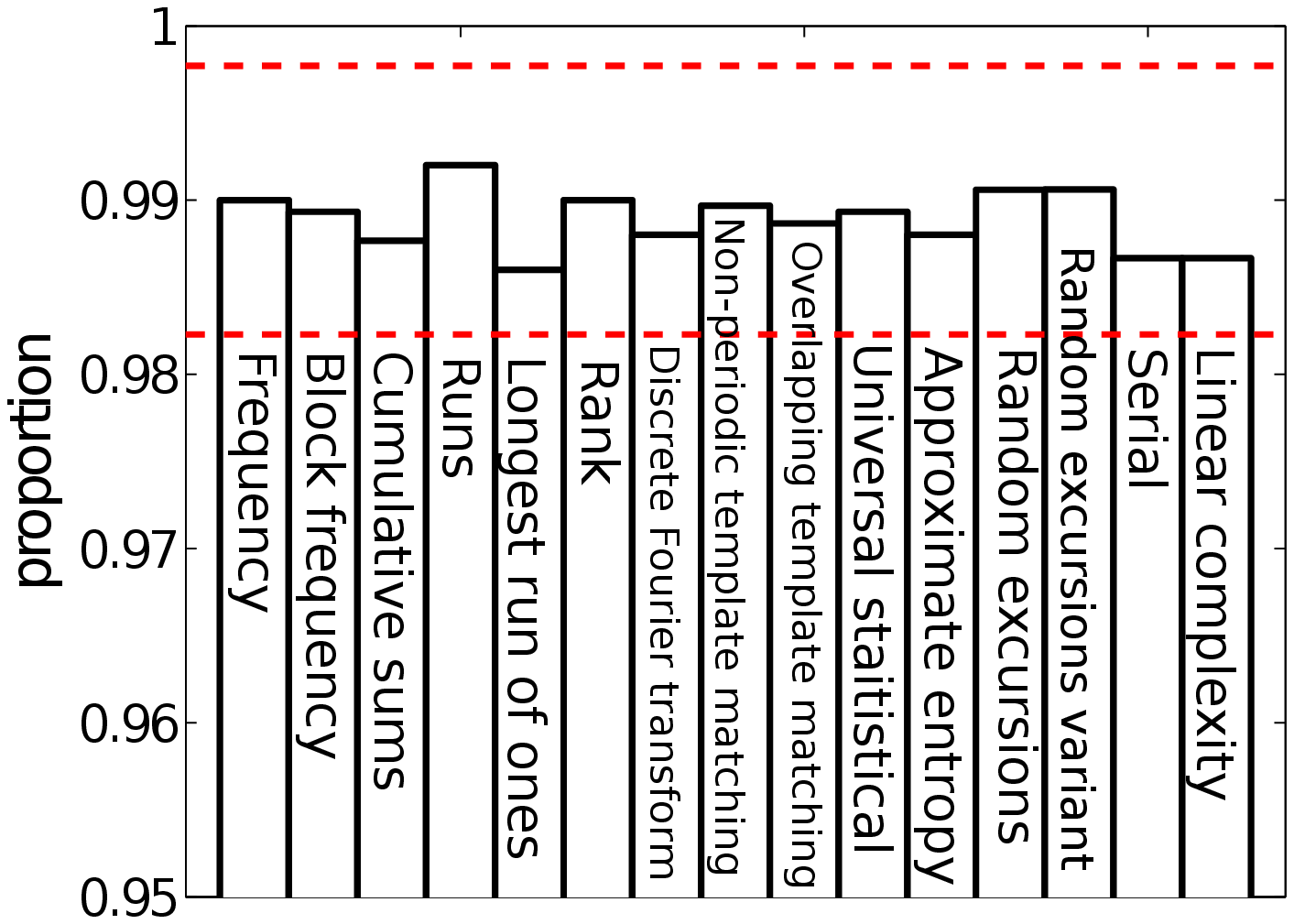}}
\subfigure[$\epsilon$-uniformity]{\label{Fig:eunif}\includegraphics[scale=0.4]{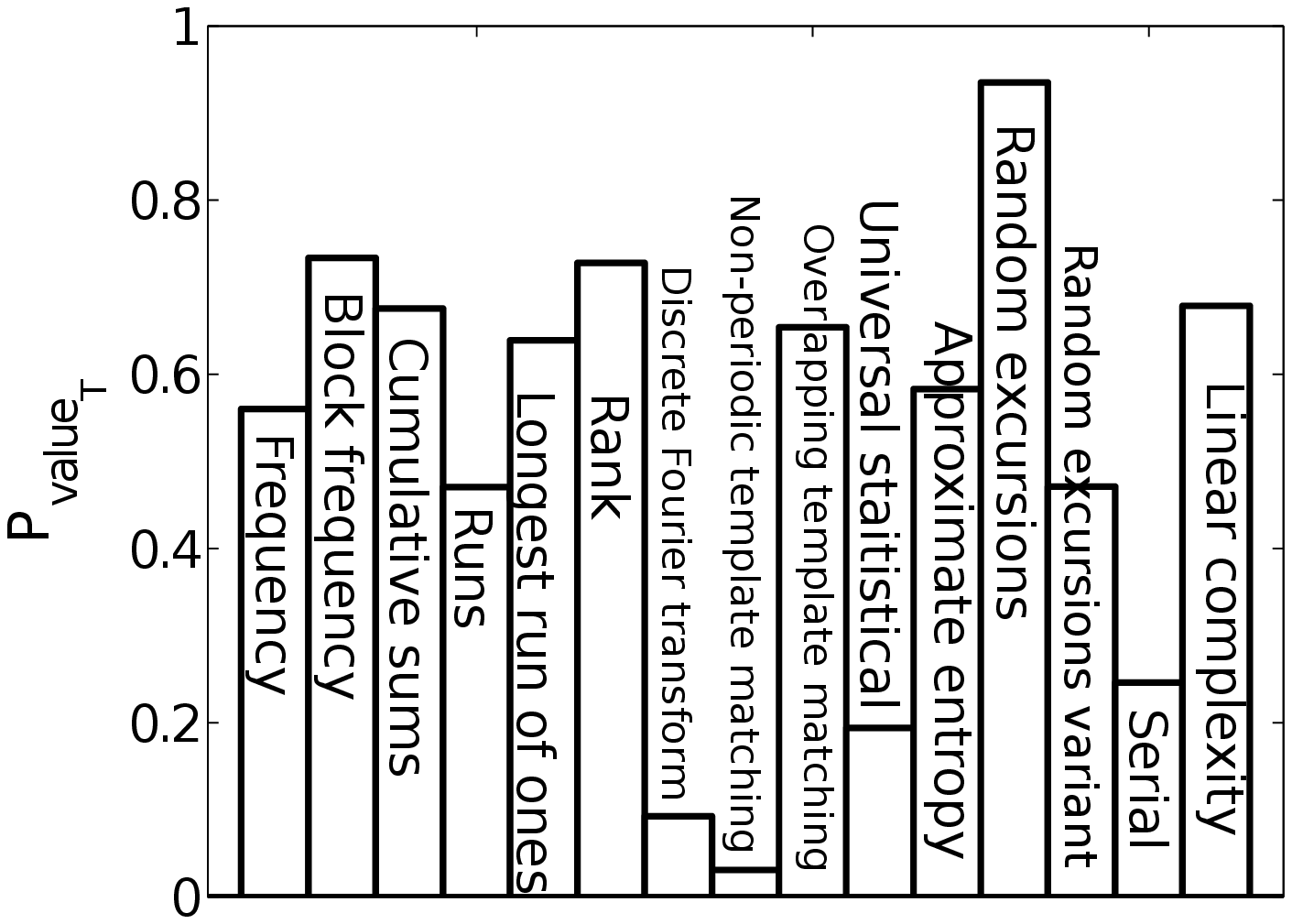}}
\caption{Summary of the results of the NIST test suite to assess randomness. Dashed lines in Fig. \ref{Fig:proportion} represent the confidence interval where the proportion of sequences accepted/rejected per test must fall in. In Fig. \ref{Fig:eunif} is plot $P_{value_T} = \incompletegamma(9/2, \chi^2/2)$ for each test. All $P_{value_T} > 10^{-4}$. The smallest one is for the non-periodic template matching test, giving a value of $P_{value_T} = 0.0926$.}
\label{Fig:NistTests}
\end{figure}

\subsection{\textbf{Statistical testing}}

The 15-test battery proposed by NIST is applied on the hashed data to assess its randomness. The significance level ($\alpha_{\textsc{sl}}$) is set at 0.01. It means that 1 in a 100 sequences is expected to be rejected even if it is produced by a fair random generator. Using suggestions in \cite{Rukhin2010}, the proportion of accepted/rejected sequences and the uniformity of the $P$-values are computed. As shown in Fig. \ref{Fig:NistTests}, all tests are passed with a sequence of $1.5$ Gbits.

In order to evaluate the results of the tests, two statistics are calculated. First, the ratio of accepted to rejected sequences is calculated and must fall within the confidence interval defined by $\big[ 1-\alpha_{\textsc{sl}} \pm 3\sqrt{(1 - \alpha_{\textsc{sl}})\alpha_{\textsc{sl}}/m} \big]$, where $m = 1500$ is the number of 1 Mbit sequences tested, see Fig. \ref{Fig:proportion}. Second, the $\epsilon$-uniformity of the $P$-values is examined. The idea is to compute $P_{value_T}$, a \textit{`$P$-value of $P$-values'}. The procedure is as follows: for each test, (i) calculate a 10-bit histogram of $P$-values, (ii) compute the $\chi^2$ defined in Eq. (\ref{Eq:XiS}) and (iii) calculate the incomplete gamma statistical function $\incompletegamma(9/2, \chi^2/2)$, which must be larger than $10^{-4}$, see Fig. \ref{Fig:eunif}.
\begin{equation}
\label{Eq:XiS}
\chi^2 = \frac{s}{10} \sum_{i=1}^{10} \Big(F_i - \frac{s}{10}\Big)^2,
\end{equation} 
where $s$ is the number of $P$-values per test and $F_i$ is the number of $P$-values in the $i$-th bin.

\section{Conclusions}
We have demonstrated a quantum random number generator (QRNG) based on interferometric detection of phase diffusion in a high bandwidth gain-switched DFB laser diode.  The method uses interferometry rather than single-photon or shot-noise-limited detection to convert microscopic quantum observables to macroscopically detectable signals, and is demonstrated with commercially available telecommunications components.  These features make it robust, low cost, compact and low power consumption. Using a 5.825 GHz excitation rate and 14-bit digitization, we observe {43} Gbps quantum randomness generation.  We compute a conservative bound on the phase diffusion between pulses and find diffusion by at least 9.45 radians rms, equivalent to complete phase randomness for all practical purposes.  The generated bit strings show no detectable correlations or deviations from uniformity, and pass all tests in the NIST 15-test battery.  We provide a new method to determine the quantum contribution to the detected random signals, and show that the quantum contribution to min-entropy grows proportional to the digitization resolution in bits, suggesting a straightforward route to even higher QRNG bit rates.  The high rate as well as the quantum nature of the randomness generation make this scheme attractive for cryptography, gambling and quantum key distribution.  

\section*{Acknowledgments}
This work was supported by the ERC under project MAMBO (Proof of Concept of PERCENT) and project AQUMET, MINECO under projects FIS2011-23520, TEC2010-14832, and Explora INTRINQRA, Galician Regional Government under projects CN2012/279 and CN2012/260 ``Consolidation of research units: AtlantTIC'' and FEDER under project Ref: UPVOV10-3E-492.  
 \\\\


\end{document}